\documentclass[year=26,pdfa]{fmcad}

\usepackage{amsmath,amssymb,amsfonts}
\usepackage[ruled,linesnumbered,noend]{algorithm2e}
\usepackage{graphicx}
\usepackage{textcomp}
\usepackage{booktabs}
\usepackage{multirow}
\usepackage{xspace}
\usepackage{url}
\usepackage{xcolor}
\usepackage{pifont}
\usepackage{tikz}
\usetikzlibrary{positioning}

\newcommand{\always}{\square}
\newcommand{\eventually}{\Diamond}
\newcommand{\Next}{\bigcirc}
\newcommand{\until}{\mathbin{\mathcal{U}}}
\newcommand{\GRone}{GR(1)\xspace}
\newcommand{\toolname}{\textsc{GR1Mine}\xspace}

\begin{document}

\title{Learning GR(1) Specifications from Traces}

\author{
  \IEEEauthorblockN{Sam Nicholas Kouteili\,\orcid{0009-0007-9713-7191}}
  \IEEEauthorblockA{Yale University\\
  New Haven, USA\\
  sam.kouteili@yale.edu}
    \and
  \IEEEauthorblockN{William Fishell}
  \IEEEauthorblockA{Columbia University\\
  New York, USA\\
  wf2322@columbia.edu}
  \and
  \IEEEauthorblockN{Mark Santolucito\,\orcid{0000-0001-8646-4364}}
  \IEEEauthorblockA{Barnard College, Columbia University\\
  New York, USA\\
  msantolu@barnard.edu}
  \and
  \IEEEauthorblockN{Ruzica Piskac\,\orcid{0000-0002-3267-0776}}
  \IEEEauthorblockA{Yale University\\
  New Haven, USA\\
  ruzica.piskac@yale.edu}
}

\maketitle


\begin{abstract}
Constrained specification mining enables the automatic discovery of desired properties from system traces. Generalized Reactivity of Rank 1, or \GRone, is a fragment of LTL with polynomial-time synthesis that natively encodes assume-guarantee properties present in most hardware and robotics domains.
In this paper, we present \toolname, a SAT-based tool for efficiently learning \GRone formulas
from examples.
We exploit the \GRone temporal skeleton to incrementally enumerate formula candidates, leveraging learnt clauses to avoid recomputation.
On the Boolean \GRone Syntech suite,
\toolname learns a realizable formula for
all 60 benchmarks over 30$\times$ faster than generic and constrained LTL mining tools. 
On non-GR(1) specifications from SYNTCOMP, \toolname is still able to recover $>$2$\times$ more realizable specifications than baselines within the timeout.

\end{abstract}

\section{Introduction}\label{sec:intro}

Generalized Reactivity of Rank~1, or \GRone, is a fragment of Linear Temporal Logic (LTL) that naturally captures a broad class of
reactive systems. 
Several real-world systems, from industrial hardware bus
protocols such as the AMBA AHB bus arbiter~\cite{BloemGJPPW07}, to robotic motion
planning~\cite{KressGazitFP09} and autonomous
controllers~\cite{MaozR21}, fall in this class of specifications.
In \GRone, a system must maintain ongoing obligations
provided the environment satisfies its own fairness
conditions~\cite{PitermanPS06}.
\GRone is particularly attractive because of its tractability for
reactive synthesis: while synthesis from general LTL specifications
is doubly exponential in the length of the
formula~\cite{PnueliR89}, \GRone synthesis is polynomial in the
size of the state space~\cite{PitermanPS06}. 

Noting this, 
a formal \GRone model of the system may not be available in practical settings. 
Consider a scenario such as the one presented in~\cite{ye2025hardwareoracle}, where a black-box hardware controller
generates execution traces.
With positive and negative traces representing correct and undesirable system behavior
(i.e. from test logs with good and bad
executions), we can
learn a specification discriminating the two.
The mined specification not only reveals underlying reactive properties of the system, but can be fed directly into
existing \GRone synthesis tools to produce correct-by-construction
controllers.
This enables a pipeline from observed behaviors to formally verified implementation, with a declarative model of the system.

Existing specification mining
tools~\cite{NeiderG18,LemieuxDB15} learn minimal LTL
formulas from positive and negative examples.
However, the mined formulas do not
distinguish between environment assumptions and system guarantees, meaning that synthesis must treat
all inputs as adversarial. This often renders the mined
specification unrealizable, even when a realizable \GRone specification exists.

Moreover, minimal formulas produced by LTL learners may fail to capture intended system behavior. 
Suppose that a developer calls an LTL miner wanting to uncover reactive properties such as $\always(a \rightarrow \eventually b)$, but instead is given a non-reactive formula $a \vee \neg b$ that minimally separates the samples. This has inspired works such as ATLAS~\cite{ZhangCGD25} that allow users to specify structural constraints on learnt formulas.
While one can encode the syntactic shape of a \GRone specification in ATLAS, constrained learners treat the \GRone skeleton as a filter on a general LTL search space rather than exploiting it
algorithmically:
scalability thus prohibitively degrades.

In this paper, we present \toolname, a SAT-based algorithm for learning \GRone formulas from traces.
\toolname fixes the \GRone  skeleton
and encodes the propositional content of each
component as a Directed Acyclic Graph (DAG).
By encoding component DAG constraints once and sharing them across enumerated candidate formulas, we exploit the enumerative search space via
incremental SAT solving.
An environment--system variable partition 
ensures that
mined formulas respect the assume--guarantee boundary and are suitable for synthesis.

We evaluate \toolname on all Boolean \GRone specifications from the
Syntech benchmarks~\cite{MaozR21}, mining a \GRone formula for all 60. 
Conversely, ATLAS in unconstrained and
\GRone-constrained configurations solves 33 and 22 formulas respectively within the 3-minute timeout. Evaluating on non-\GRone benchmarks from SYNTCOMP, \toolname mines formulas for 38/60 benchmarks, with ATLAS[LTL] recovering 18 and ATLAS[GR1] only recovering 1.
The formulas mined by ATLAS[LTL] are either unrealizable or trivial, yielding 1-state controllers.
In all, our contributions are as follows:
\begin{enumerate}
    \item We develop a novel algorithm for learning \GRone specifications from examples, leveraging learnt clauses over enumerated formulas.
    \item We implement \toolname, an efficient \GRone learner that discriminates positive and negative traces.
    \item We benchmark our tool against unconstrained and constrained LTL solvers. We find that we tractably solve the most benchmarks and produce rich, realizable formulas.
\end{enumerate}



\section{Preliminaries}\label{sec:prelims}

\subsection{Linear Temporal Logic}

We consider LTL over a set of atomic propositions $\mathit{AP} = \{x_0, \ldots, x_{n-1}\}$.
Formulas are built from the grammar
\[
  \varphi ::= x \mid \lnot\varphi \mid \varphi \land \varphi \mid
  \Next\varphi \mid \varphi \until \varphi
\]
where $x \in \mathit{AP}$, and the derived operators \emph{eventually}
$\eventually\varphi \equiv \mathit{true} \until \varphi$ and \emph{always}
$\always\varphi \equiv \lnot\eventually\lnot\varphi$.
Semantics are defined over infinite words $w \in (2^{\mathit{AP}})^\omega$ in the standard way~\cite{Pnueli77}.

\subsection{Generalized Reactivity of Rank 1}\label{sec:gr1}

A GR(1) specification~\cite{PitermanPS06} over $\mathit{AP}$ has the form
\begin{equation}\label{eq:gr1}
  \underbrace{(\varphi_I^e \land \always \varphi_S^e \land
  \textstyle\bigwedge_{i=1}^{m} \always\eventually J_i)}_{\text{assumptions } \mathcal{A}}
  \;\rightarrow\;
  \underbrace{(\varphi_I^s \land \always \varphi_S^s \land
  \textstyle\bigwedge_{j=1}^{k} \always\eventually G_j)}_{\text{guarantees } \mathcal{G}}
\end{equation}
where $\varphi_I^e, \varphi_I^s$ are propositional \emph{initial conditions},
$\varphi_S^e, \varphi_S^s$ are propositional \emph{safety conditions},
and $J_1, \ldots, J_m$ and $G_1, \ldots, G_k$ are propositional
\emph{justice (liveness) conditions}.
Every component is a propositional formula over $\mathit{AP}$ (i.e., built from $\land$, $\lor$, $\lnot$ and variables only); the temporal operators $\always$, $\eventually$ appear only in the fixed positions shown above.

\subsection{Lasso Traces}\label{sec:lasso}

An \emph{ultimately periodic word} (lasso trace) is a pair $(u, v)$ with $u \in (2^{\mathit{AP}})^*$ and $v \in (2^{\mathit{AP}})^+$, representing the infinite word $u \cdot v^\omega$. We write $\sigma \models \varphi$ when the infinite word $\sigma = u \cdot v^\omega$ satisfies $\varphi$ under LTL semantics~\cite{Pnueli77}.
We store a lasso trace as a finite vector $\sigma = \sigma[0] \cdots \sigma[L{-}1]$ together with a \emph{lasso start} index $\ell$, so that $\sigma[\ell..L{-}1]$ is the repeating loop.
We write $\mathit{loop}(\sigma) = \{\ell, \ldots, L{-}1\}$ for the set of loop positions.

A lasso trace $\sigma$ satisfies $\always\eventually J$ iff $J$
holds at some position in $\mathit{loop}(\sigma)$; it satisfies
$\always \varphi_S$ iff $\varphi_S$ holds at every position, and it
satisfies $\varphi_I$ iff $\varphi_I$ holds at position~0.
The full formula~\eqref{eq:gr1} is an implication: $\sigma$ satisfies
it iff either some assumption fails, or all guarantees hold.

\subsection{Reactive Synthesis}

Reactive synthesis is the problem of automatically constructing a system implementation (transducer) that satisfies a given temporal specification for all possible environment behaviors~\cite{PnueliR89}. The input is a specification $\varphi$ over environment and system variables; the output is a strategy (or controller) that chooses system outputs in response to environment inputs such that $\varphi$ is guaranteed. \GRone is the largest fragment of LTL for which reactive synthesis
is known to be polynomial: the three-nested fixed-point algorithm of Piterman~et~al.~\cite{PitermanPS06} runs in $O(n^3 m k)$ on a game graph with $n$ states, compared to 2EXPTIME-complete general LTL synthesis (note that the game graph is exponential in the number of variables)~\cite{PnueliR89}.

In reactive synthesis, propositions are partitioned into
\emph{environment} (input) variables $\mathit{AP}_{env}$ and
\emph{system} (output) variables $\mathit{AP}_{sys}$.
Assumptions constrain the environment and guarantees constrain the
system.
We adopt the standard contract on how components may reference
this partition---a common sufficient (though not necessary)
restriction adopted for tractable synthesis, rather than a
realizability requirement: justice conditions $J_i$ use only $\mathit{AP}_{env}$,
guarantee conditions $G_j$ use only $\mathit{AP}_{sys}$,
initial conditions are split ($\varphi_I^e$ over $\mathit{AP}_{env}$,
$\varphi_I^s$ over $\mathit{AP}_{sys}$), and safety conditions may
reference all current-state variables but restrict
\emph{next-state} (primed) variables to their respective
side.

\subsection{Specification Mining}

Given a set of positive traces $P$ and negative traces $N$,
the \emph{specification mining problem} is to find a formula $\varphi$ that separates $P$ and $N$. Formally, that is to say learn $\varphi$ such that $\sigma \models \varphi$ for all $\sigma \in P$ and
$\varsigma \not\models \varphi$ for all $\varsigma \in N$.


\section{Algorithm}\label{sec:algo}

\begin{figure}[t]
\centering
\begin{tikzpicture}[scale=0.9, every node/.style={transform shape},
  skel/.style={font=\scriptsize, draw, rounded corners=1.5pt,
               minimum height=0.32cm, inner sep=1.5pt, fill=gray!10},
  skeldash/.style={font=\scriptsize, draw, dashed, rounded corners=1.5pt,
               minimum height=0.32cm, inner sep=1.5pt, fill=white, text=gray},
  temporal/.style={font=\scriptsize},
  dagnode/.style={draw, circle, minimum size=0.26cm, inner sep=0pt,
                  font=\tiny, fill=blue!8},
  dagdash/.style={draw, circle, minimum size=0.26cm, inner sep=0pt,
                  font=\tiny, dashed, fill=white, text=gray},
   dagdot/.style={draw, circle, minimum size=0.26cm, inner sep=0pt,
                  font=\tiny, dotted, fill=white, text=gray},
  conn/.style={-, very thin, blue!50},
  arr/.style={->, >=stealth, very thin},
  darr/.style={->, >=stealth, very thin, dashed, gray},
  lbl/.style={draw=none, font=\small\bfseries},
  note/.style={draw=none, font=\tiny, text=black!55},
  uarr/.style={->, >=stealth, very thin, black!35},
]
\def\colA{0.6}
\def\colB{3.0}
\def\colC{5.7}
\def\dotsX{7.5}
\def\rA{0}       
\def\rB{-1.55}   
\def\rC{-3.65}   

\node[lbl] at (-0.85, \rA) {$D\!=\!1$};

\node[note] at (\colA+0.5, \rA+0.6) {\scriptsize$m\!=\!1,k\!=\!1$};
\node[temporal] at (\colA-0.45, \rA+0.28) {\tiny$\always\eventually$};
\node[skel] (j1a) at (\colA-0.45, \rA) {$J_1$};
\node[temporal] at (\colA+0.05, \rA) {\tiny$\to$};
\node[temporal] at (\colA+0.5, \rA+0.28) {\tiny$\always\eventually$};
\node[skel] (g1a) at (\colA+0.5, \rA) {$G_1$};
\node[dagnode] at (\colA-0.45, \rA-0.38) {$?$};
\node[dagnode] at (\colA+0.5, \rA-0.38) {$?$};
\draw[conn] (j1a.south) -- ++(0,-0.13);
\draw[conn] (g1a.south) -- ++(0,-0.13);

\draw[uarr] (\colA+1.05, \rA) -- node[above, font=\tiny, text=black!35]
  {\textsc{unsat}} ++(0.35, 0);

\node[note] at (\colB+0.5, \rA+0.6) {\scriptsize$m\!=\!1,k\!=\!2$};
\node[temporal] at (\colB-0.45, \rA+0.28) {\tiny$\always\eventually$};
\node[skel] (j1b) at (\colB-0.45, \rA) {$J_1$};
\node[temporal] at (\colB+0.05, \rA) {\tiny$\to$};
\node[temporal] at (\colB+0.45, \rA+0.28) {\tiny$\always\eventually$};
\node[skel] (g1b) at (\colB+0.45, \rA) {$G_1$};
\node[temporal] at (\colB+0.95, \rA+0.28) {\tiny$\always\eventually$};
\node[skeldash] (g2b) at (\colB+0.95, \rA) {$G_2$};
\node[dagnode] at (\colB-0.45, \rA-0.38) {$?$};
\node[dagnode] at (\colB+0.45, \rA-0.38) {$?$};
\node[dagdash] at (\colB+0.95, \rA-0.38) {$?$};
\draw[conn] (j1b.south) -- ++(0,-0.13);
\draw[conn] (g1b.south) -- ++(0,-0.13);
\draw[conn, dashed, gray] (g2b.south) -- ++(0,-0.13);

\draw[uarr] (\colB+1.4, \rA) -- node[above, font=\tiny, text=black!35]
  {\textsc{unsat}} ++(0.35, 0);

\node[note] at (\colC+0.45, \rA+0.6) {\scriptsize$m\!=\!2,k\!=\!1$};
\node[temporal] at (\colC-0.55, \rA+0.28) {\tiny$\always\eventually$};
\node[skel] (j1c) at (\colC-0.55, \rA) {$J_1$};
\node[temporal] at (\colC+0.0, \rA+0.28) {\tiny$\always\eventually$};
\node[skeldash] (j2c) at (\colC+0.0, \rA) {$J_2$};
\node[temporal] at (\colC+0.5, \rA) {\tiny$\to$};
\node[temporal] at (\colC+1.0, \rA+0.28) {\tiny$\always\eventually$};
\node[skel] (g1c) at (\colC+1.0, \rA) {$G_1$};
\node[dagnode] at (\colC-0.55, \rA-0.38) {$?$};
\node[dagdash] at (\colC+0.0, \rA-0.38) {$?$};
\node[dagnode] at (\colC+1.0, \rA-0.38) {$?$};
\draw[conn] (j1c.south) -- ++(0,-0.13);
\draw[conn, dashed, gray] (j2c.south) -- ++(0,-0.13);
\draw[conn] (g1c.south) -- ++(0,-0.13);

\node[note] at (\dotsX, \rA) {$\cdots$};

\draw[arr, black!35] (-0.55, \rA-0.7) --
  node[left, font=\tiny, text=black!35] {\rotatebox{90}{\textsc{unsat}}}
  (-0.55, \rB+0.45);

\node[lbl] at (-0.85, \rB) {$D\!=\!2$};
\def\treeS{0.3}  

\node[note] at (\colA+0.5, \rB+0.6) {\scriptsize$m\!=\!1,k\!=\!1$};
\node[temporal] at (\colA-0.45, \rB+0.28) {\tiny$\always\eventually$};
\node[skel] (j1d) at (\colA-0.45, \rB) {$J_1$};
\node[temporal] at (\colA+0.05, \rB) {\tiny$\to$};
\node[temporal] at (\colA+0.5, \rB+0.28) {\tiny$\always\eventually$};
\node[skel] (g1d) at (\colA+0.5, \rB) {$G_1$};
\node[dagnode] (d2j1r) at (\colA-0.45, \rB-0.4) {$?$};
\node[dagdash] (d2j1l) at (\colA-0.45-\treeS, \rB-0.85) {$?$};
\node[dagdash] (d2j1rr) at (\colA-0.45+\treeS, \rB-0.85) {$?$};
\draw[arr] (d2j1r.south west) -- (d2j1l.north);
\draw[arr] (d2j1r.south east) -- (d2j1rr.north);
\draw[conn] (j1d.south) -- (d2j1r.north);
\node[dagnode] (d2g1r) at (\colA+0.5, \rB-0.4) {$?$};
\node[dagdash] (d2g1l) at (\colA+0.5-\treeS, \rB-0.85) {$?$};
\node[dagdash] (d2g1rr) at (\colA+0.5+\treeS, \rB-0.85) {$?$};
\draw[arr] (d2g1r.south west) -- (d2g1l.north);
\draw[arr] (d2g1r.south east) -- (d2g1rr.north);
\draw[conn] (g1d.south) -- (d2g1r.north);

\draw[uarr] (\colA+1.05, \rB) -- node[above, font=\tiny, text=black!35]
  {\textsc{unsat}} ++(0.35, 0);

\node[note] at (\colB+0.5, \rB+0.6) {\scriptsize$m\!=\!1,k\!=\!2$};
\node[temporal] at (\colB-0.45, \rB+0.28) {\tiny$\always\eventually$};
\node[skel] (j1e) at (\colB-0.45, \rB) {$J_1$};
\node[temporal] at (\colB+0.05, \rB) {\tiny$\to$};
\node[temporal] at (\colB+0.45, \rB+0.28) {\tiny$\always\eventually$};
\node[skel] (g1e) at (\colB+0.45, \rB) {$G_1$};
\node[temporal] at (\colB+0.95, \rB+0.28) {\tiny$\always\eventually$};
\node[skeldash] (g2e) at (\colB+0.95, \rB) {$G_2$};
\node[dagnode] (d2j1er) at (\colB-0.45, \rB-0.4) {$?$};
\node[dagdash] (d2j1el) at (\colB-0.7, \rB-0.85) {$?$};
\node[dagdash] (d2j1err) at (\colB-0.2, \rB-0.85) {$?$};
\draw[arr] (d2j1er.south west) -- (d2j1el.north);
\draw[arr] (d2j1er.south east) -- (d2j1err.north);
\draw[conn] (j1e.south) -- (d2j1er.north);
\node[dagnode] (d2g1er) at (\colB+0.45, \rB-0.4) {$?$};
\node[dagdash] (d2g1el) at (\colB+0.22, \rB-0.85) {$?$};
\node[dagdash] (d2g1ec) at (\colB+0.51, \rB-0.85) {$?$};
\draw[arr] (d2g1er.south west) -- (d2g1el.north);
\draw[arr] (d2g1er.south) -- (d2g1ec.north);
\draw[conn] (g1e.south) -- (d2g1er.north);
\node[dagdash] (d2g2er) at (\colB+0.95, \rB-0.4) {$?$};
\node[dagdash] (d2g2ec) at (\colB+0.9, \rB-0.85) {$?$};
\node[dagdash] (d2g2err) at (\colB+1.15, \rB-0.85) {$?$};
\draw[darr] (d2g2er.south) -- (d2g2ec.north);
\draw[darr] (d2g2er.south east) -- (d2g2err.north);
\draw[conn, dashed, gray] (g2e.south) -- (d2g2er.north);

\draw[uarr] (\colB+1.4, \rB) -- node[above, font=\tiny, text=black!35]
  {\textsc{unsat}} ++(0.35, 0);

\node[note] at (\colC+0.45, \rB+0.6) {\scriptsize$m\!=\!2,k\!=\!1$};
\node[temporal] at (\colC-0.55, \rB+0.28) {\tiny$\always\eventually$};
\node[skel] (j1f2) at (\colC-0.55, \rB) {$J_1$};
\node[temporal] at (\colC+0.0, \rB+0.28) {\tiny$\always\eventually$};
\node[skeldash] (j2f2) at (\colC+0.0, \rB) {$J_2$};
\node[temporal] at (\colC+0.5, \rB) {\tiny$\to$};
\node[temporal] at (\colC+1.0, \rB+0.28) {\tiny$\always\eventually$};
\node[skel] (g1f2) at (\colC+1.0, \rB) {$G_1$};
\node[dagnode] (d2j1fr) at (\colC-0.55, \rB-0.4) {$?$};
\node[dagdash] (d2j1fl) at (\colC-0.7, \rB-0.85) {$?$};
\node[dagdash] (d2j1frr) at (\colC-0.4, \rB-0.85) {$?$};
\draw[arr] (d2j1fr.south) -- (d2j1fl.north);
\draw[arr] (d2j1fr.south east) -- (d2j1frr.north);
\draw[conn] (j1f2.south) -- (d2j1fr.north);
\node[dagdash] (d2j2fr) at (\colC+0.0, \rB-0.4) {$?$};
\node[dagdash] (d2j2fl) at (\colC-0.05, \rB-0.85) {$?$};
\node[dagdash] (d2j2frr) at (\colC+0.2, \rB-0.85) {$?$};
\draw[darr] (d2j2fr.south) -- (d2j2fl.north);
\draw[darr] (d2j2fr.south east) -- (d2j2frr.north);
\draw[conn, dashed, gray] (j2f2.south) -- (d2j2fr.north);
\node[dagnode] (d2g1fr) at (\colC+1.0, \rB-0.4) {$?$};
\node[dagdash] (d2g1fl) at (\colC+0.75, \rB-0.85) {$?$};
\node[dagdash] (d2g1frr) at (\colC+1.25, \rB-0.85) {$?$};
\draw[arr] (d2g1fr.south west) -- (d2g1fl.north);
\draw[arr] (d2g1fr.south east) -- (d2g1frr.north);
\draw[conn] (g1f2.south) -- (d2g1fr.north);

\node[note] at (\dotsX, \rB) {$\cdots$};

\draw[arr, black!35] (-0.55, \rB-0.85-0.2) --
  node[left, font=\tiny, text=black!35] {\rotatebox{90}{\textsc{unsat}}}
  (-0.55, \rC+0.45);

\node[lbl] at (-0.85, \rC) {$D\!=\!3$};
\def\treeD{0.45}  
\def\treeSm{0.22} 

\node[note] at (\colA+0.5, \rC+0.6) {\scriptsize$m\!=\!1,k\!=\!1$};
\node[temporal] at (\colA-0.45, \rC+0.28) {\tiny$\always\eventually$};
\node[skel] (j1f) at (\colA-0.45, \rC) {$J_1$};
\node[temporal] at (\colA+0.05, \rC) {\tiny$\to$};
\node[temporal] at (\colA+0.5, \rC+0.28) {\tiny$\always\eventually$};
\node[skel] (g1f) at (\colA+0.5, \rC) {$G_1$};
\node[dagnode] (t1jr) at (\colA-0.45, \rC-0.4) {$?$};
\node[dagdash] (t1ji) at (\colA-0.45-\treeS, \rC-0.4-\treeD) {$?$};
\node[dagdash] (t1jc) at (\colA-0.45+\treeS, \rC-0.4-\treeD) {$?$};
\node[dagdash] (t1jll) at (\colA-0.45-\treeS-\treeSm, \rC-0.4-2*\treeD) {$?$};
\node[dagdash] (t1jlr) at (\colA-0.45-\treeS+\treeSm, \rC-0.4-2*\treeD) {$?$};
\draw[arr] (t1jr.south west) -- (t1ji.north);
\draw[arr] (t1jr.south east) -- (t1jc.north);
\draw[arr] (t1ji.south west) -- (t1jll.north);
\draw[arr] (t1ji.south east) -- (t1jlr.north);
\draw[conn] (j1f.south) -- (t1jr.north);
\node[dagnode] (t1gr) at (\colA+0.5, \rC-0.4) {$?$};
\node[dagdash] (t1gi) at (\colA+0.5-\treeS, \rC-0.4-\treeD) {$?$};
\node[dagdash] (t1gc) at (\colA+0.5+\treeS, \rC-0.4-\treeD) {$?$};
\node[dagdash] (t1gll) at (\colA+0.5-\treeS-\treeSm, \rC-0.4-2*\treeD) {$?$};
\node[dagdash] (t1glr) at (\colA+0.5-\treeS+\treeSm, \rC-0.4-2*\treeD) {$?$};
\draw[arr] (t1gr.south west) -- (t1gi.north);
\draw[arr] (t1gr.south east) -- (t1gc.north);
\draw[arr] (t1gi.south west) -- (t1gll.north);
\draw[arr] (t1gi.south east) -- (t1glr.north);
\draw[conn] (g1f.south) -- (t1gr.north);

\draw[uarr] (\colA+1.05, \rC) -- node[above, font=\tiny, text=black!35]
  {\textsc{unsat}} ++(0.35, 0);

\node[note] at (\colB+0.5, \rC+0.6) {\scriptsize$m\!=\!1,k\!=\!2$};
\node[temporal] at (\colB-0.45, \rC+0.28) {\tiny$\always\eventually$};
\node[skel] (j1g) at (\colB-0.45, \rC) {$J_1$};
\node[temporal] at (\colB+0.05, \rC) {\tiny$\to$};
\node[temporal] at (\colB+0.45, \rC+0.28) {\tiny$\always\eventually$};
\node[skel] (g1g) at (\colB+0.45, \rC) {$G_1$};
\node[temporal] at (\colB+0.95, \rC+0.28) {\tiny$\always\eventually$};
\node[skeldash] (g2g) at (\colB+0.95, \rC) {$G_2$};
\node[dagnode] (t2jr) at (\colB-0.45, \rC-0.4) {$?$};
\node[dagdash] (t2ji) at (\colB-0.7, \rC-0.4-\treeD) {$?$};
\node[dagdash] (t2jc) at (\colB-0.2, \rC-0.4-\treeD) {$?$};
\node[dagdash] (t2jll) at (\colB-0.7-\treeSm, \rC-0.4-2*\treeD) {$?$};
\node[dagdash] (t2jlr) at (\colB-0.7+\treeSm, \rC-0.4-2*\treeD) {$?$};
\draw[arr] (t2jr.south west) -- (t2ji.north);
\draw[arr] (t2jr.south east) -- (t2jc.north);
\draw[arr] (t2ji.south west) -- (t2jll.north);
\draw[arr] (t2ji.south east) -- (t2jlr.north);
\draw[conn] (j1g.south) -- (t2jr.north);
\node[dagnode] (t2g1r) at (\colB+0.45, \rC-0.4) {$?$};
\node[dagdash] (t2g1i) at (\colB+0.22, \rC-0.4-\treeD) {$?$};
\node[dagdash] (t2g1c) at (\colB+0.51, \rC-0.4-\treeD) {$?$};
\node[dagdash] (t2g1ll) at (\colB+0.22-\treeSm, \rC-0.4-2*\treeD) {$?$};
\node[dagdash] (t2g1lr) at (\colB+0.14+\treeSm, \rC-0.4-2*\treeD) {$?$};
\draw[arr] (t2g1r.south west) -- (t2g1i.north);
\draw[arr] (t2g1r.south) -- (t2g1c.north);
\draw[arr] (t2g1i.south west) -- (t2g1ll.north);
\draw[arr] (t2g1i.south east) -- (t2g1lr.north);
\draw[conn] (g1g.south) -- (t2g1r.north);
\node[dagdash] (t2g2r) at (\colB+0.95, \rC-0.4) {$?$};
\node[dagdash] (t2g2i) at (\colB+0.82, \rC-0.4-\treeD) {$?$};
\node[dagdash] (t2g2c) at (\colB+1.15, \rC-0.4-\treeD) {$?$};
\node[dagdash] (t2g2ll) at (\colB+0.9-\treeSm, \rC-0.4-2*\treeD) {$?$};
\node[dagdash] (t2g2lr) at (\colB+0.82+\treeSm, \rC-0.4-2*\treeD) {$?$};
\draw[darr] (t2g2r.south west) -- (t2g2i.north);
\draw[darr] (t2g2r.south east) -- (t2g2c.north);
\draw[darr] (t2g2i.south west) -- (t2g2ll.north);
\draw[darr] (t2g2i.south east) -- (t2g2lr.north);
\draw[conn, dashed, gray] (g2g.south) -- (t2g2r.north);

\draw[uarr] (\colB+1.4, \rC) -- node[above, font=\tiny, text=black!35]
  {\textsc{unsat}} ++(0.35, 0);

\node[note] at (\colC+0.45, \rC+0.6) {\scriptsize$m\!=\!2,k\!=\!1$};
\node[temporal] at (\colC-0.55, \rC+0.28) {\tiny$\always\eventually$};
\node[skel] (j1h) at (\colC-0.55, \rC) {$J_1$};
\node[temporal] at (\colC+0.0, \rC+0.28) {\tiny$\always\eventually$};
\node[skeldash] (j2h) at (\colC+0.0, \rC) {$J_2$};
\node[temporal] at (\colC+0.5, \rC) {\tiny$\to$};
\node[temporal] at (\colC+1.0, \rC+0.28) {\tiny$\always\eventually$};
\node[skel] (g1h) at (\colC+1.0, \rC) {$G_1$};
\node[dagnode] (t3j1r) at (\colC-0.55, \rC-0.4) {$?$};
\node[dagdash] (t3j1i) at (\colC-0.7, \rC-0.4-\treeD) {$?$};
\node[dagdash] (t3j1c) at (\colC-0.4, \rC-0.4-\treeD) {$?$};
\node[dagdash] (t3j1ll) at (\colC-0.7-\treeSm, \rC-0.4-2*\treeD) {$?$};
\node[dagdash] (t3j1lr) at (\colC-0.8+\treeSm, \rC-0.4-2*\treeD) {$?$};
\draw[arr] (t3j1r.south) -- (t3j1i.north);
\draw[arr] (t3j1r.south east) -- (t3j1c.north);
\draw[arr] (t3j1i.south west) -- (t3j1ll.north);
\draw[arr] (t3j1i.south east) -- (t3j1lr.north);
\draw[conn] (j1h.south) -- (t3j1r.north);
\node[dagdash] (t3j2r) at (\colC+0.0, \rC-0.4) {$?$};
\node[dagdash] (t3j2i) at (\colC-0.1, \rC-0.4-\treeD) {$?$};
\node[dagdash] (t3j2c) at (\colC+0.2, \rC-0.4-\treeD) {$?$};
\node[dagdash] (t3j2ll) at (\colC-0.001-\treeSm, \rC-0.4-2*\treeD) {$?$};
\node[dagdash] (t3j2lr) at (\colC-0.1+\treeSm, \rC-0.4-2*\treeD) {$?$};
\draw[darr] (t3j2r.south) -- (t3j2i.north);
\draw[darr] (t3j2r.south east) -- (t3j2c.north);
\draw[darr] (t3j2i.south west) -- (t3j2ll.north);
\draw[darr] (t3j2i.south east) -- (t3j2lr.north);
\draw[conn, dashed, gray] (j2h.south) -- (t3j2r.north);
\node[dagnode] (t3g1r) at (\colC+1.0, \rC-0.4) {$?$};
\node[dagdash] (t3g1i) at (\colC+0.82, \rC-0.4-\treeD) {$?$};
\node[dagdash] (t3g1c) at (\colC+1.18, \rC-0.4-\treeD) {$?$};
\node[dagdash] (t3g1ll) at (\colC+0.82-\treeSm, \rC-0.4-2*\treeD) {$?$};
\node[dagdash] (t3g1lr) at (\colC+0.82+\treeSm, \rC-0.4-2*\treeD) {$?$};
\draw[arr] (t3g1r.south west) -- (t3g1i.north);
\draw[arr] (t3g1r.south east) -- (t3g1c.north);
\draw[arr] (t3g1i.south west) -- (t3g1ll.north);
\draw[arr] (t3g1i.south east) -- (t3g1lr.north);
\draw[conn] (g1h.south) -- (t3g1r.north);

\node[note] at (\dotsX, \rC) {$\cdots$};

\end{tikzpicture}
\caption{Two-dimensional search in \toolname.
\textbf{Horizontal}: template enumeration over $(m,k)$ via
push/pop.
\textbf{Vertical}: number of literals $D$.
$D\!=\!1$: single literal.
$D\!=\!2$: two literals, one operator ($\lnot x_0 \land x_1$, etc.).
$D\!=\!3$: three literals, two operators.
Solid = current; dashed = newly added.}\label{fig:search}
\end{figure}
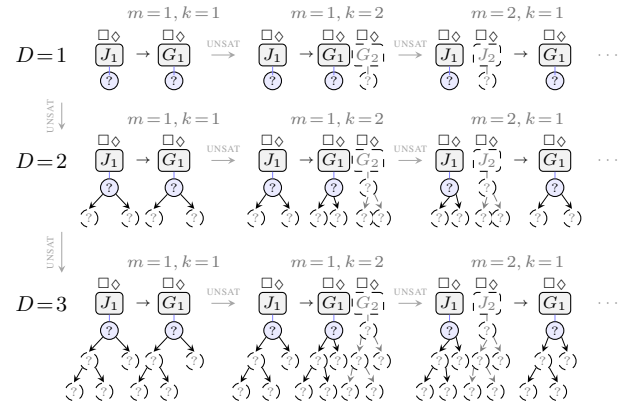

We now present \toolname, a SAT-based procedure for mining \GRone
specifications from lasso traces.
\toolname performs a two-dimensional search: the outer loop
increases the number of literals $D$ (vertical axis),
controlling how complex each component can be
($D{=}1$: a single variable, e.g.\ $x_0$ or $\lnot x_0$;
$D{=}2$: two literals, e.g.\ $\lnot x_0 \land x_1$; $D{=}3$: three literals, e.g. $(\lnot x_0 \vee x_1) \land x_2$);
the inner loop enumerates template configurations $(m,k,F)$
(horizontal axis), varying the number of justice and guarantee
conditions.
For each combination, \toolname encodes the mining problem as a SAT
instance and checks satisfiability, returning the first solution
found. Figure~\ref{fig:search} illustrates \toolname's search strategy.

The \GRone template~\eqref{eq:gr1} is parameterized by the number
of justice conditions $m$, guarantees $k$, and a subset
$F \subseteq \{\mathit{init}, \mathit{safe}\}$ indicating which
optional component types to include: $\mathit{init} \in F$ means
the template includes initial-condition components
($\mathit{init}_\mathit{env}$ on the assumption side,
$\mathit{init}_\mathit{sys}$ on the guarantee side), and
$\mathit{safe} \in F$ means safety components are included.
Since the correct template is unknown, \toolname searches $(m, k, F)
\in \{1,\ldots,M\} \times \{1,\ldots,K\} \times
2^{\{\mathit{init},\,\mathit{safe}\}}$
in non-decreasing order of total component count
$m + k + 2|F|$.

We pre-allocate a set of \emph{component identifiers}
$\mathcal{C}$,
one for each slot in the maximal template.
Each $c \in \mathcal{C}$ names an \emph{unknown} propositional
formula to be determined by the SAT solver: for instance,
$\mathit{init}_\mathit{env}$ names the environment initial condition
whose content $\varphi_I^e$ is unknown.
Our per-component encoding builds on the DAG-based SAT
encoding of Neider and Gavran~\cite{NeiderG18}, restricted to
propositional operators.
The SAT variables $x^c, n^c, l^c, r^c$ encode the syntax DAG for~$c$
(Section~\ref{sec:component-dags}),
$y^c$ evaluates it on traces
(Section~\ref{sec:semantics}),
and $h^c$ captures whether it holds under its temporal wrapper
(Section~\ref{sec:temporal}).
The \textsc{EncodeAll} subroutine in
Algorithm~\ref{alg:main} bundles these steps for all
components at a given depth.

The key observation enabling incremental solving is that all
of these constraints are \emph{independent} of the template
assignment, with only the consistency
constraints~(\eqref{eq:pos}--\eqref{eq:neg}), asserted in
lines~\ref{ln:pos} and~\ref{ln:neg} of Algorithm~\ref{alg:main},
changing.
For each depth $D$, we encode all $c \in \mathcal{C}$ once,
then iterate over template configurations.
Each configuration partitions the active components into
assumption-side ($J_i$, $\mathit{init}_\mathit{env}$, $\mathit{safe}_\mathit{env}$) and
guarantee-side ($G_j$, $\mathit{init}_\mathit{sys}$, $\mathit{safe}_\mathit{sys}$),
asserts the consistency constraints~\eqref{eq:pos}--\eqref{eq:neg}
inside a \textsc{Push}/\textsc{Pop} scope, and queries the solver.
\textsc{Push} saves a checkpoint on the solver's
constraint stack; \textsc{Pop} retracts all constraints added since
the checkpoint, restoring the solver to its prior state.
Learned clauses persist across checkpoints, so information
gained from one UNSAT configuration prunes subsequent
configurations.


Algorithm~\ref{alg:main} gives the complete procedure;
the remaining subsections detail the four families of constraints
added by \textsc{EncodeAll}: DAG structure
(Section~\ref{sec:component-dags}), propositional semantics
(Section~\ref{sec:semantics}), temporal holds
(Section~\ref{sec:temporal}), and trace consistency
(Section~\ref{sec:consistency}).

\begin{algorithm}[t]
\caption{\toolname}\label{alg:main}
\scriptsize
\DontPrintSemicolon
\SetKwInOut{Input}{Input}
\SetKwInOut{Output}{Output}
\SetKwFunction{Push}{Push}
\SetKwFunction{Pop}{Pop}
\SetKwFunction{Sat}{Sat}
\SetKwFunction{add}{add}
\SetKwFunction{Reconstruct}{Reconstruct}
\SetKwFunction{Encode}{EncodeAll}
\SetKwProg{Proc}{Procedure}{}{}
\Input{Traces ($P$, $N$),
  partition $(\mathit{AP}_\mathit{env}, \mathit{AP}_\mathit{sys})$,
  bounds $M$, $K$}
\Output{\GRone formula separating $P$ from $N$}
\BlankLine
\Proc{\Encode{$\mathcal{S}, \mathcal{C}, D$}}{
  \ForEach{$c \in \mathcal{C}$}{
    Init $x^c_{i,p}, n^c_i$
      $\forall\, i \!<\! D$,
      $p \!\in\! \mathit{Vars}(c)$ \tcp*{leaves, Sec.~\ref{sec:component-dags}}
    Init $x^c_{i,\lambda}$, $l^c_{i,j}$, $r^c_{i,j}$
      $\forall\, D \!\le\! i \!<\! 2D{-}1$,
      $\lambda \!\in\! \{\land,\lor\}$,
      $j \!<\! i$ \tcp*{internal nodes}
    Init $y^c_{i,\sigma,t}$
      $\forall\, i \!<\! 2D{-}1$, $\sigma \!\in\! P \!\cup\! N$, $t \!<\! |\sigma|$ \tcp*{Sec.~\ref{sec:semantics}}
    Init $h^c_\sigma$
      $\forall\, \sigma \!\in\! P \!\cup\! N$ \tcp*{Sec.~\ref{sec:temporal}}
    $\mathcal{S}$.\add{one-hot labels, arity, acyclicity for
      $x^c, l^c, r^c$}\;
    $\mathcal{S}$.\add{propositional semantics of $y^c$ via
      $x^c, l^c, r^c$}\;
    $\mathcal{S}$.\add{$h^c_\sigma$ via
      Eqs.~\eqref{eq:init}--\eqref{eq:guarantee}}\;
      
  }
}
\noindent\textbf{end procedure}\;
\BlankLine
$\mathcal{S} \gets \textsc{SatSolver}.\textsc{new}()$\;
$\mathcal{C} \gets \{J_1,\ldots,J_M,\, G_1,\ldots,G_K,\,
  \mathit{init}_\mathit{env},\, \mathit{init}_\mathit{sys},\,
  \mathit{safe}_\mathit{env},\, \mathit{safe}_\mathit{sys}\}$\;
\For{$D = 1, 2, 3, \ldots$}{
  \Encode{$\mathcal{S}, \mathcal{C}, D$}
    \tcp*{permanent constraints}
  \For{$(m, k, F) \in
    \{1,\!\ldots\!,M\} {\times} \{1,\!\ldots\!,K\} {\times}
    2^{\{\mathit{init},\,\mathit{safe}\}}$}{
    $\mathcal{A} \gets \{J_1,\!\ldots\!,J_m\}$;\;
    $\mathcal{G} \gets \{G_1,\!\ldots\!,G_k\}$\;
    \lIf{$\mathit{init} \in F$}{$\mathcal{A} \gets \mathcal{A}
      \cup \{\mathit{init}_\mathit{env}\}$;\,
      $\mathcal{G} \gets \mathcal{G}
      \cup \{\mathit{init}_\mathit{sys}\}$}
    \lIf{$\mathit{safe} \in F$}{$\mathcal{A} \gets \mathcal{A}
      \cup \{\mathit{safe}_\mathit{env}\}$;\,
      $\mathcal{G} \gets \mathcal{G}
      \cup \{\mathit{safe}_\mathit{sys}\}$}
    $\mathcal{S}$.\Push{}
      \tcp*{ckpt: below is retractable}
    \ForEach{$\sigma \in P$}{
      $\mathcal{S}$.\add{$
        \bigwedge_{a \in \mathcal{A}} h^a_\sigma \Rightarrow
        \bigwedge_{g \in \mathcal{G}} h^g_\sigma$}
        \tcp*{Eq.~\eqref{eq:pos}}\label{ln:pos}
    }
    \ForEach{$\sigma \in N$}{
      $\mathcal{S}$.\add{$
        \bigwedge_{a \in \mathcal{A}} h^a_\sigma \land
        \lnot(\bigwedge_{g \in \mathcal{G}} h^g_\sigma)$} \tcp*{Eq.~\eqref{eq:neg}}\label{ln:neg}
    }
    \lIf{$\mathcal{S}$.\Sat{}}{
      \Return \Reconstruct{model, $\mathcal{A}, \mathcal{G}$}}
    $\mathcal{S}$.\Pop{}
      \tcp*{retract consistency constraints}
  }
}
\end{algorithm}

\subsection{Component DAGs}\label{sec:component-dags}

Each component $c$ of the \GRone template is an unknown
propositional formula over literals combined with $\land$ and $\lor$.
We represent it as a \emph{literal-leaves tree} with $D$
leaf nodes and $D{-}1$ internal nodes, for $2D{-}1$ nodes total
(a single leaf when $D{=}1$).
Nodes $0, \ldots, D{-}1$ are \emph{leaves} (one literal each);
nodes $D, \ldots, 2D{-}2$ are \emph{internal} (one binary operator
each); and the root is node $2D{-}2$ for $D \ge 1$.
For a given~$D$, all DAG variables are created once by
\textsc{EncodeAll} (Alg.~\ref{alg:main}) and persist across
template configurations; only consistency constraints
(Sec.~\ref{sec:consistency}) change.

The tree is described by four families of SAT variables:
\begin{itemize}
\item \emph{Variable selection}\; $x^c_{i,p}$: ``leaf $i$ selects
    atomic proposition~$p$,'' where $p \in \mathit{Vars}(c)$,
    the set of propositions permitted in component $c$ by the
    environment--system partition.
    Exactly one $p$ is selected per leaf.
    For internal nodes, $x^c_{i,\lambda}$ with
    $\lambda \in \{\land, \lor\}$ selects the binary operator.
\item \emph{Polarity}\; $n^c_i$: ``leaf $i$ is negated.''
    When true, the leaf represents $\lnot p$ rather than~$p$.
\item \emph{Wiring}\; $l^c_{i,j}$, $r^c_{i,j}$: ``the left
    (right) child of internal node $i$ is node $j$,'' for
    $D \le i < 2D{-}1$, $0 \le j < i$.
    Each internal node has exactly one left and one right child.
\item \emph{Evaluation}\; $y^c_{i,\sigma,t}$: ``node $i$
    evaluates to true at position $t$ of trace~$\sigma$.''
    These variables 
    are
    constrained by the propositional semantics
    (Section~\ref{sec:semantics}).
\end{itemize}


\subsection{Propositional Semantics}\label{sec:semantics}

The $y$ variables link each tree to the traces.
For a leaf~$i$ selecting proposition $x_p$ with polarity
flag~$n^c_i$, the evaluation is:
$y^c_{i,\sigma,t} \Leftrightarrow \sigma[t][p] \texttt{ XOR } n^c_i$.
For primed leaves $x_p'$ in safety components:
$y^c_{i,\sigma,t} \Leftrightarrow \sigma[\mathit{next}(t)][p]
\texttt{ XOR } n^c_i$.
For internal nodes, the evaluation is determined by
the node's operator and children:
\begin{align}
x^c_{i,\land} \land l^c_{i,a} \land r^c_{i,b} &\;\Rightarrow\;
  \bigl(y^c_{i,\sigma,t} \Leftrightarrow
  y^c_{a,\sigma,t} \land y^c_{b,\sigma,t}\bigr) \\
x^c_{i,\lor} \land l^c_{i,a} \land r^c_{i,b} &\;\Rightarrow\;
  \bigl(y^c_{i,\sigma,t} \Leftrightarrow
  y^c_{a,\sigma,t} \lor y^c_{b,\sigma,t}\bigr)
\end{align}
These are asserted for every trace $\sigma$ and position $t$.


\subsection{Temporal Holds}\label{sec:temporal}

Rather than searching over temporal operators, we hardcode
the temporal semantics of each component's wrapper.
A single Boolean variable $h^c_\sigma$ captures whether component
$c$ holds on trace~$\sigma$, defined in terms of the root node's
$y$ values.
Let $\rho$ denote the root node index $2D{-}2$.
For a trace $\sigma$ of length $L$ with lasso start $\ell$:
\begin{align}
  h^{\mathit{init}}_\sigma &\;\Leftrightarrow\;
    y^{\mathit{init}}_{\rho,\,\sigma,\,0}
    \label{eq:init}\\[2pt]
  h^{\mathit{safety}}_\sigma &\;\Leftrightarrow\;
    \textstyle\bigwedge_{t=0}^{L-1}
    y^{\mathit{safety}}_{\rho,\,\sigma,\,t}
    \label{eq:safety}\\[2pt]
  h^{J_i}_\sigma &\;\Leftrightarrow\;
    \textstyle\bigvee_{t \in \mathit{loop}(\sigma)}
    y^{J_i}_{\rho,\,\sigma,\,t}
    \label{eq:justice}\\[2pt]
  h^{G_j}_\sigma &\;\Leftrightarrow\;
    \textstyle\bigvee_{t \in \mathit{loop}(\sigma)}
    y^{G_j}_{\rho,\,\sigma,\,t}
    \label{eq:guarantee}
\end{align}


\subsection{Consistency Constraints}\label{sec:consistency}

The consistency constraints are the only constraints that change
as the algorithm iterates over configurations $(m, k, F)$
(Alg.~\ref{alg:main}, lines~\ref{ln:pos} and~\ref{ln:neg}).
They enforce the assume--guarantee semantics of \GRone on the given traces, specifically expressing that every positive trace that satisfies all
assumptions must also satisfy all guarantees, and every negative
trace must satisfy all assumptions while violating at least one
guarantee. 
While iterating through template configurations, if the current configuration is unsatisfiable, the solver retracts the present consistency constraints and tries the next configuration.
Formally, for each positive trace $\sigma \in P$:
\begin{equation}\label{eq:pos}
\textstyle\bigwedge_{a \in \mathcal{A}} h^a_\sigma
\;\Rightarrow\;
\textstyle\bigwedge_{g \in \mathcal{G}} h^g_\sigma
\end{equation}
For each negative trace $\sigma \in N$:
\begin{equation}\label{eq:neg}
\textstyle\bigwedge_{a \in \mathcal{A}} h^a_\sigma
\;\;\land\;\;
\lnot\bigl(\textstyle\bigwedge_{g \in \mathcal{G}} h^g_\sigma\bigr)
\end{equation}




\section{Evaluation}\label{sec:eval}

We evaluate \toolname on a suite of specifications drawn from
real reactive-synthesis benchmarks, comparing against the
state-of-the-art constrained LTL learner
ATLAS~\cite{ZhangCGD25} in two configurations. 
All experiments run on a single core of an Apple~M1 Pro ARM machine
with a 3-minute timeout.
Our evaluation aims to address two research questions:

\textbf{RQ1)} How effectively does \toolname learn specifications from traces generated by \GRone systems? What about non-\GRone systems? 

\textbf{RQ2)} How meaningful are learnt specifications for synthesis? Are mined formulas realizable, and if so, are synthesized controllers non-trivial?


\subsection{Benchmarks}\label{sec:benchmarks}

We draw our \GRone benchmarks from the \textsc{Syntech} repository of the Spectra
project~\cite{MaozR21}.
We consider all Syntech benchmarks from the repository that have not been sugared with syntactic extensions (enumerated types, array indexing). The \emph{boolean-only} subsets end up being the
\texttt{bloemDebugging}~\cite{BloemJPPS12} and
\texttt{cimattiAnalyzing}~\cite{CimattiGMT20} directories,
which yield 60 parseable specifications covering hardware-protocol families.



To evaluate \toolname on specifications outside the \GRone fragment,
we sample 60 TLSF benchmarks from the \textsc{SyntComp} synthesis
competition~\cite{JacobsBBKP17}.
We consider arbitrary LTL safety/liveness specifications not of the \GRone form, making them a stress test
for whether \toolname's template-restricted mining still produces
useful results when the source specification is not \GRone-shaped.

For each benchmark, we generate 20~positive and 20~negative lasso
traces.
Half the traces in each class are produced by a generator that constructs assumption-satisfying traces via Z3~\cite{z3}:
positive traces satisfy both assumptions and guarantees, while
negative traces satisfy the assumptions but violate at least one
guarantee.
The remaining traces are generated by \emph{random sampling} of lasso
traces, often violating the assumptions. We pad all traces to be length 6, finding that ATLAS does not properly handle inputs with varied trace lengths.
We settled on length~6 after preliminary runs with trace lengths 4--8, where longer traces did not empirically capture more complex portions of the original specifications; formula complexity and trace count, rather than trace length, are the dominant runtime bottlenecks.
With this mix, we garner rich and varied traces that respect and violate both the assumptions and the guarantees.

\subsection{Baselines}\label{sec:baselines}

We compare \toolname against ATLAS~\cite{ZhangCGD25}, a recent
constrained LTL learner that encodes the mining problem as a
MaxSAT instance via the Alloy modeling language and
the AlloyMax solver.
ATLAS accepts user-specified structural constraints in the form of
Alloy predicates, allowing it to restrict the search space.
We evaluate against two configurations.

\begin{itemize}
\item \textbf{ATLAS[LTL]}: unconstrained LTL mining over the full
  operator set $\{\always, \eventually, \Next, \until, \land, \lor,
  \lnot, \rightarrow\}$.
  Mined formulas are synthesized via
\texttt{ltlsynt}~\cite{MichaudC18}.
\item \textbf{ATLAS[GR1]}: \GRone template
  and the environment--system variable partition encoded as Alloy predicates.
  As configured, the Alloy fact fixes the
  template to
  $\always\eventually J \rightarrow \always\eventually G$,
  with no safety or initial-condition components: ATLAS does not
  support optional template structure, so covering the full \GRone
  template~\eqref{eq:gr1} would require enumerating a separate
  encoding for each subset of optional components.
\end{itemize}

\noindent
ATLAS[GR1] and \toolname formulas are synthesized via the \GRone synthesis tool \texttt{slugs}~\cite{EhlersR16}.

\subsection{Results}\label{sec:results}

Table~\ref{tab:results} summarizes the results grouped by
variable count.
Detailed per-benchmark results with mined formulas are given
in the accompanying artifact.

\begin{table}[t]
\centering
\caption{Results across Syntech and Syntcomp benchmarks. \emph{Slv}: solved within 3\,min. \emph{U}: number of unrealizable mined formula (red, shown when nonzero). \emph{Time}: cumulative wallclock (seconds).}\label{tab:results}
\vspace{2pt}
\scriptsize
\setlength{\tabcolsep}{3pt}
\begin{tabular}{@{}l r r r r r r r r@{}}
\toprule
& &
  \multicolumn{2}{c}{\toolname} &
  \multicolumn{2}{c}{ATLAS[LTL]} &
  \multicolumn{2}{c}{ATLAS[GR1]} \\
\cmidrule(lr){3-4}\cmidrule(lr){5-6}\cmidrule(lr){7-8}
Family & $n$ &
  Slv(U) & Time(s) &
  Slv(U) & Time(s) &
  Slv(U) & Time(s) \\
\midrule
\multicolumn{8}{l}{\textbf{SYNTECH}} \\
\midrule
A2+AMBA (2) & 8 & \textbf{8} & 356s & 1\textbf{\textcolor{red}{(1)}} & 1272s & 0 & 1440s \\
A4+AMBA (4) & 8 & \textbf{8} & 10s & 8\textbf{\textcolor{red}{(1)}} & 266s & 6 & 640s \\
A5+AMBA (5) & 8 & \textbf{8} & 15s & \textbf{8} & 475s & 4 & 1050s \\
AMBA & 6 & \textbf{6} & 128s & 2 & 753s & 2 & 912s \\
G100 & 5 & \textbf{5} & 29s & 0 & 900s & 0 & 900s \\
G20+GenBuf (20) & 8 & \textbf{8} & 13s & 4\textbf{\textcolor{red}{(1)}} & 1420s & 0 & 1440s \\
G5+GenBuf (5) & 8 & \textbf{8} & 14s & 7 & 262s & \textbf{8} & 784s \\
GenBuf & 9 & \textbf{9} & 30s & 3 & 1212s & 2 & 1413s \\
\midrule
\emph{SYNTECH subtotal} & \emph{60} & \emph{60} & \emph{595s} & \emph{33\textbf{\textcolor{red}{(3)}}} & \emph{6561s} & \emph{22} & \emph{8579s} \\
\midrule
\multicolumn{8}{l}{\textbf{SYNTCOMP}} \\
\midrule
gui\_glue\_code\_synthesis & 2 & \textbf{2} & 1s & \textbf{2} & 3s & 1 & 185s \\
lily & 9 & 5 & 746s & 1\textbf{\textcolor{red}{(1)}} & 1528s & 0 & 1620s \\
ltl2dba & 7 & 0 & 1260s & 0 & 1260s & 0 & 1260s \\
ltl2dpa & 7 & 5 & 557s & 0 & 1260s & 0 & 1260s \\
sweap & 7 & \textbf{7} & 6s & 7\textbf{\textcolor{red}{(7)}} & 130s & 0 & 1260s \\
tsl\_paper & 7 & 4 & 559s & 4 & 603s & 0 & 1260s \\
tsl\_smart\_home\_jarvis & 21 & 15 & 1667s & 4\textbf{\textcolor{red}{(1)}} & 3240s & 0 & 3780s \\
\midrule
\emph{SYNTCOMP subtotal} & \emph{60} & \emph{38} & \emph{4796s} & \emph{18\textbf{\textcolor{red}{(9)}}} & \emph{8025s} & \emph{1} & \emph{10625s} \\
\midrule
\textbf{Total} & \textbf{120} & \textbf{98} & \textbf{5391s} & \textbf{51\textbf{\textcolor{red}{(12)}}} & \textbf{14586s} & \textbf{23} & \textbf{19204s} \\
\bottomrule
\end{tabular}
\end{table}

\textbf{RQ1} \textbf{How effectively does \toolname learn from traces?}

\toolname solves all 60 Spectra benchmarks (100\%) within the timeout. ATLAS[LTL] solves 33/60 (55\%), while ATLAS[GR1] solves only 22/60 (37\%). ATLAS[GR1] times out on all A2+AMBA, G100, and G20+GenBuf benchmarks: 
encoding the GR(1) template as Alloy predicates atop full LTL semantics proves prohibitively expensive at these sizes. On A5+AMBA and G5+GenBuf, all three tools solve most benchmarks, but \toolname is significantly faster (avg. $\leq$1s vs. 17–29s for ATLAS[LTL] and 60–68s for ATLAS[GR1]). 


On the 60 SYNTCOMP LTL benchmarks, \toolname mines a GR(1) formula on 38/60 (63\%). ATLAS[LTL] mines a formula on 18/60 (30\%), and ATLAS[GR1] mines only 1/60. Under the same template constraint as \toolname, the Alloy/MaxSAT encoding is essentially unable to make progress. The 22 SYNTCOMP benchmarks where \toolname returns no specification are precisely those whose ground truth contains explicitly non-GR(1) patterns (e.g.\ persistence/recurrence patterns over nested $\Next$).

\begin{figure}[t]
\centering
\includegraphics[width=\columnwidth]{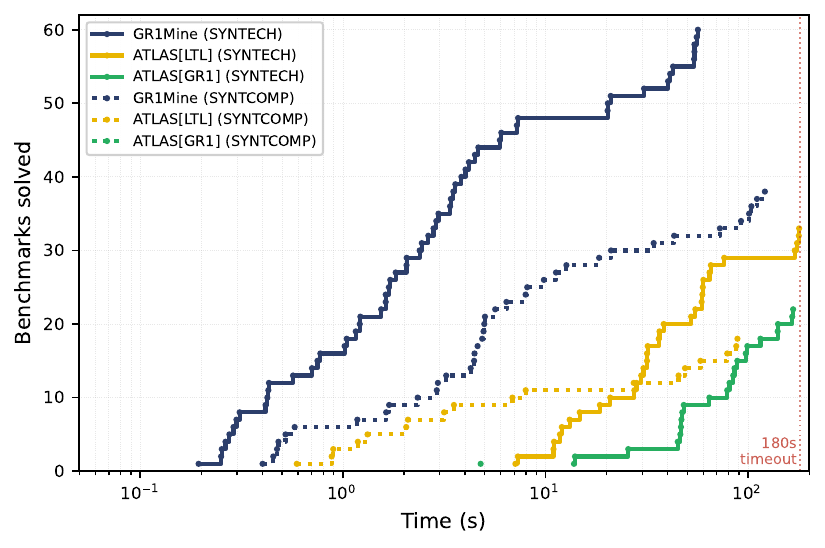}
\caption{Cactus plot: cumulative benchmarks solved vs.\ wall-clock
time.}\label{fig:cactus}
\end{figure}

\begin{figure}[t]
\centering
\includegraphics[width=\columnwidth]{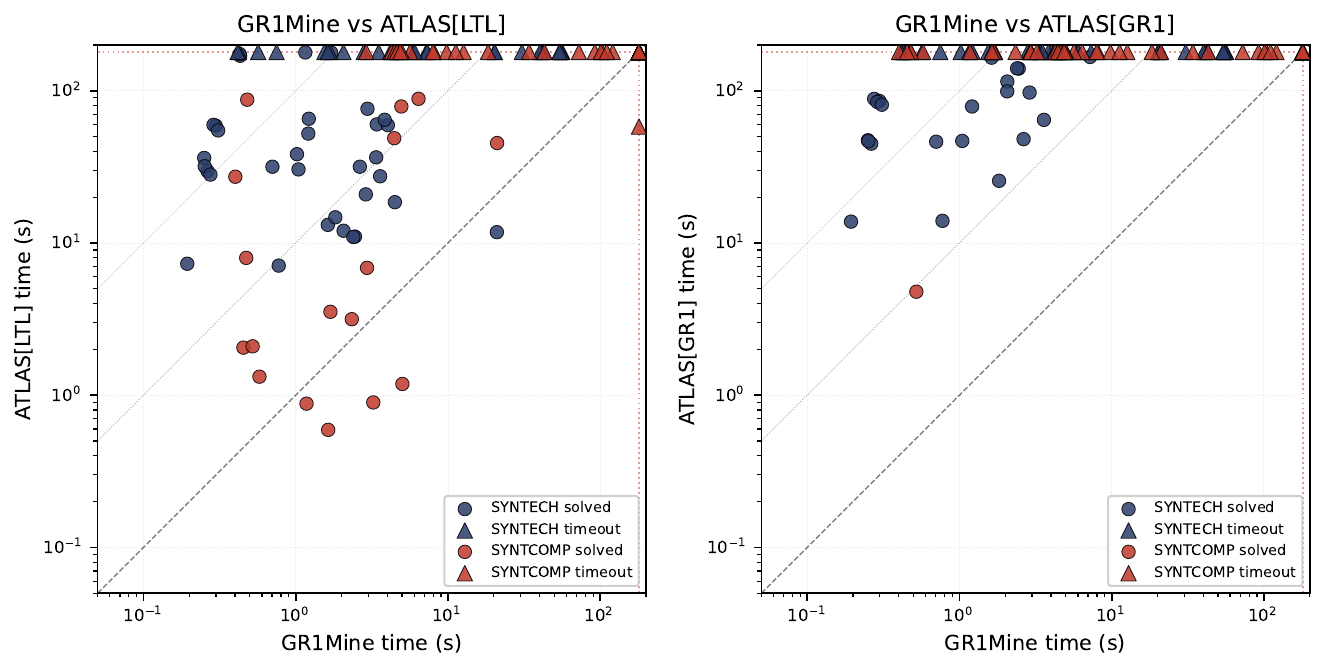}
\caption{Scatter plots comparing mining time (log scale).
Points above the diagonal indicate \toolname is faster.
Triangles denote timeouts.}\label{fig:scatter}
\end{figure}

Figure~\ref{fig:cactus} shows a cactus plot of cumulative benchmarks
solved as a function of wall-clock time, with solid lines for the Syntech benchmarks and dotted lines for SYNTCOMP. Here, we see that \toolname solves both benchmark sets roughly an Order of Magnitude (OoM) faster than ATLAS[LTL] and $\sim$2 OoM faster than ATLAS[GR1]. Figure~\ref{fig:scatter} provides a more granular insight into per-benchmark solve times. On every benchmark solved by both, \toolname is faster than ATLAS[GR1]; ATLAS[LTL] beats \toolname on 6/120 benchmarks.

\textbf{RQ2} \textbf{How suitable are learnt specifications for synthesis?} 

Alongside the ability to capture a wide range of reactive systems, \GRone specifications are particularly suitable for synthesis as they natively distinguish assumptions from guarantees.  
Table~\ref{tab:results} shows that all 98~\toolname formulas and all
23~ATLAS[GR1] formulas are realizable, while only 39 of 51~ATLAS[LTL]
formulas are (76\%; 3~unrealizable on Syntech, 9~on SYNTCOMP). Both
\GRone-constrained miners benefit from the env/sys partition baked into
the template.
ATLAS[LTL] mines a flat LTL separator over the full vocabulary,
drawing no such distinction. As such, the synthesizer then treats the whole
formula as a guarantee against an adversarial environment, and any
clause that obliges the system to make an \emph{environment} variable
hold is immediately unrealizable.

For example, on \texttt{gen\_buf\_w\_guar\_fair\_20\_genbuf} ATLAS[LTL]
returns $\eventually\, x_{16}$, where $x_{16}$ is \texttt{StoB\_REQ16}, an
environment request line from sender~16. \texttt{ltlsynt} treats this as
a guarantee, asking the controller to ensure an input it cannot drive
eventually goes high. An adversarial environment holds
\texttt{StoB\_REQ16} low forever, so no controller exists. 
Every
ATLAS[LTL] unrealizability we observed has this property.
\toolname avoids this by construction: \texttt{StoB\_REQ16} can
only land inside a justice \emph{assumption}
$\always\eventually\,\texttt{StoB\_REQ16}$, never a guarantee.
The remaining (realizable) ATLAS[LTL] formulas are trivial
separators such as $\eventually x_i$ or simple disjunctions.


\section{Related Work}\label{sec:related}

\paragraph{Learning Specifications from Traces}
Flie~\cite{NeiderG18} was a seminal work for learning the smallest LTL formula separating
positive from negative lasso traces by SAT encodings over syntax
DAGs. Our propositional encoding builds on this framework but
eliminates temporal operator search by fixing the \GRone skeleton. Tools such as Scarlet~\cite{RahaRFV22} and Bolt~\cite{BathieFMMV26} have been developed for the finite-trace LTL$_f$ setting, where traces are not lasso-shaped and temporal operator semantics change.
Fijalkow and Lagarde~\cite{FijalkowL21} establish complexity bounds
for passive LTL learning, showing NP-hardness even for restricted fragments that are weaker than GR(1), e.g. LTL(X, $\land$).
All of these approaches mine monolithic temporal formulas without
an environment-system boundary.

\paragraph{Constrained Specification Mining}
Texada~\cite{LemieuxDB15} was among the first to introduce
user-specified property templates into specification mining.
LTLSketcher~\cite{LutzNR23} extends this idea to sketch
completion: given a partial LTL formula with holes, it fills them
from positive and negative examples using a SAT encoding.
ATLAS~\cite{ZhangCGD25} further generalizes structural constraints
via first-order relational logic in Alloy, reducing constrained LTL
learning to MaxSAT.
One can encode \GRone structure as an ATLAS constraint, but the underlying encoding retains
full LTL semantics, in practice leading to intractable search.


\paragraph{\GRone Learning}
Orthogonally, work on \GRone assumption
mining~\cite{ChatterjeeHJ08,LiDS11, ye2025hardwareoracle} automatically
refines environment assumptions for a \emph{given} set of system
guarantees, while ~\cite{KonighoferHB13} localizes errors in existing
\GRone specifications and \cite{georgescu2026adaptivegr1specificationrepair} repairs them.
These approaches assume that part of the specification is already
known; \toolname mines the complete assume--guarantee formula from
traces alone.


\section{Conclusion}\label{sec:conclusion}

We presented \toolname, a SAT-based algorithm for mining \GRone
specifications from lasso traces.
By fixing the temporal skeleton of the \GRone template and encoding
only propositional content, \toolname eliminates the need to search
over temporal operators and produces formulas that are directly
suitable for reactive synthesis.
An environment--system variable partition ensures that mined
specifications respect the assume--guarantee boundary, and
incremental SAT solving enables efficient enumeration over template
configurations.

Across 120~benchmarks drawn from Syntech and SYNTCOMP, \toolname solves 98 within a 3-minute timeout, and
all 98 mined specifications are realizable. Compared to the other
\GRone-constrained miner, ATLAS[GR1] (23/120 solved), \toolname is
consistently an order of magnitude faster. Compared to unconstrained LTL mining via
ATLAS[LTL] (51/120 solved), \toolname is faster on all but five small
($\le 8$~variable) benchmarks while additionally guaranteeing that
every mined formula is realizable. ATLAS[LTL] returns flat LTL separators that omit the
env/sys partition and are unrealizable in 24\% of its successful
runs.


\bibliographystyle{IEEEtran}
\bibliography{references}

\end{document}